\documentstyle[aps,twocolumn]{revtex}
\begin{document}
\draft
\title{A reply to ``Quantum Clock Synchronization''}
\author{E.A. Burt, C.R. Ekstrom and T.B. Swanson}
\address{U.S. Naval Observatory, 3450 Massachusetts AVE NW, 
Washington, DC  20392}
\date{Received July 11, 2000}
\maketitle

\begin{abstract}
Recently a protocol for Quantum Clock Synchronization (QCS) of  
remote clocks using quantum 
entanglement was proposed by Jozsa {\it et al}.  This method has the goal 
of eliminating the random noise present in classical synchronization 
techniques.  However, as stated QCS depends on the two members of 
each entangled pair undergoing the same unitary evolution even 
while being transported to different locations.  This 
is essentially equivalent to a perfect Eddington Slow Clock Transfer 
protocol and thus, not an improvement over classical techniques.  We 
will discuss this and suggest ways in which QCS may still be used. 
\end{abstract}

In a preprint entitled, ``Quantum Clock Synchronization Based on 
Shared Prior Entanglement'' (hereafter referred to as QCSP) 
recently posted to the LANL quantum physics 
archive \cite{refJozsa}, the 
authors outline a technique for using quantum non-locality to 
synchronize remote clocks without 
the usual errors inherent in classical time transfer techniques.  It 
is our view that, as stated, this scheme has several fundamental 
limitations that will prevent it from achieving clock synchronization 
better than can be accomplished using classical techniques.  It is 
worth noting at the outset that our objections are not based on the 
fact that these limitations are hard to overcome, but that they are 
inherently the same limitations found in classical synchronization 
methods.  To overcome them means that the problem is already solved 
classically, without resorting to quantum mechanical methods.

While the possibility of using 
quantum non-locality to synchronize clocks is intriguing, it would 
appear that any such scheme must be independent of {\it how} 
information is transmitted classically.  In the case of the protocol 
outlined in QCSP, information is 
being transmitted classically in two ways.  First, through the usual 
classical channel that accompanies entanglement experiments and 
second, the relative phase between two partners in an entangled pair
is transmitted with the entangled particles 
themselves as they are transported from one site to another.
Since this phase will necessarily be shifted by the classical 
transportation process, such a dependence on classical conveyance of 
this information will 
make this QCS protocol equivalent to existing classical methods such as 
the Eddington Slow Clock Transport (ESCT) protocol \cite{refESCT}.  

In ESCT two local clocks are synchronized and one of them is moved 
``slowly'' to a remote site where it can remain or where it can transfer the 
synchronization with the first clock to a third remote clock.  
``Slowly'' is defined as a velocity such that velocity-dependent 
perturbations to the clock are lower than any other noise in the 
problem.  In clock jargon this is sometimes called a ``perfect clock 
trip''.
In comparison, protocols for 
Quantum Cryptography \cite{refEkert}, which are similar to QCS, are not 
sensitive to the relative phase of the entangled particles as they are 
transmitted.

The authors start with a two-level system consisting of states 
$|0\rangle$ and $|1\rangle$ and energy eigenvalues $E_{0}$ and $E_{1}$ 
with corresponding frequency $\Omega = {1 \over \hbar} (E_{1} - E_{0})$.  
A dual basis $|pos\rangle = {1 \over \sqrt{2}}(|0\rangle + |1\rangle)$ 
and $|neg\rangle = {1 \over \sqrt{2}}(|0\rangle - |1\rangle)$ is then 
defined along with  the 
``Hadamard transform'', {\it H}, that maps 
$|0\rangle$ into $|pos\rangle$ and $|1\rangle$ into $|neg\rangle$.  
The claim is made that two consecutive applications of {\it 
H} separated by time, {\it t}, is equivalent to a Ramsey temporal 
interferometer.  While the time evolution of the states after the 
first application of {\it H} in QCSP eq. 1 is essentially correct (up to 
an overall 
time-evolving phase), the expression for the probability of measuring 
the system in one state or the other in eq. 2 is not correct in the 
context of the Ramsey method as it is applied to atomic clocks 
\cite{refRamsey}.  
In a Ramsey interferometer with interrogation on 
resonance, the probability of measuring the system in state 
$|0\rangle$ or $|1\rangle$ is independent of the time between the two 
pulses (the ``Ramsey time'').  

The problem lies in the fact that the authors have not included a key 
part of the Ramsey interferometer:  the relative phase between the 
interrogation oscillator and the precession of the system.  Later this 
phase ambiguity is addressed and we will come back to it as well.

Even though {\it H} may not correspond exactly to a Ramsey 
interferometer, it is likely that it can be realized and that it would
produce the desired 
result 
(QCSP eq. 2).  Let us assume that this is true and that the 
result of the measurements described is as stated in eq. 2.  The 
authors then state that two separated observers, Alice and Bob 
(hereafter referred to as {\it A} and {\it B}), 
share an ensemble of 
singlet states of the form, $|\psi^{-}\rangle = {1 \over \sqrt{2}} 
(|0\rangle_{A}|1\rangle_{B} - |1\rangle_{A}|0\rangle_{B})$.  
The statement is made that, ``\ldots this singlet state 
\ldots does not evolve in time {\it provided A and B undergo 
identical unitary evolutions\ldots}'' (emphasis added).  It is not 
clear from the paper whether the authors intend these unitary 
transformations to {\it include} the transportation of the entangled 
particles.  If transportation {\it is} included then the statement 
that {\it A} and {\it B} undergo identical unitary evolutions is 
equivalent to the statement that {\it A} and {\it B}
are able to build perfect clocks and transport them perfectly.  Thus, 
they have effectively solved time synchronization completely and with 
only classical methods thereby making QCS unnecessary.  

On the other 
hand, if they are not including the transportation process in the 
unitary transformations, then they still have  problems.  First, they 
are  ignoring the 
phase shift introduced by transporting one of the particles in an 
entangled pair. This phase would 
usually be caused by Doppler shifts and electromagnetic field 
perturbations.  While these can be minimized, the essential point is that they 
are of the same type whether one is referring to an entangled pair or 
an atomic clock.  Therefore, the authors seem to be implicitly 
assuming a perfect ESCT.  If one 
has perfect ESCT, then one doesn't need QCS,
at least to the extent that transporting a clock is no more 
difficult than transporting ensembles of entangled particles.

The second problem is that {\it A} and {\it B} almost certainly 
do not undergo identical unitary 
evolutions.  To do so would mean that they have perfect control over 
at least their 
environments and therefore can build perfectly stable clocks.  If 
they have perfect clocks then they have already achieved frequency 
synchronization without QCS simply by virtue of having built their respective 
clocks.  Even if one assumes that perfect clocks could be built, the 
resultant frequency synchronization could not be translated, in general, 
into time synchronization by QCS because of the transport 
problems already discussed.

The authors next define the QCS protocol:  1) an ensemble of entangled particles 
called ``pre-clocks'' is created in the $|pos\rangle,|neg\rangle$ basis; 
2) {\it A} starts the clocks by measuring her half of the ensemble in that 
basis, thereby also determining which members in her ensemble are of 
type $|pos\rangle$ (type I) and which are of type $|neg\rangle$ (type 
II); 3) {\it A} communicates type I/II information to {\it B} over a classical 
channel; 4) {\it B} selects out type II (in phase with {\it A}'s type I) particles 
and measures them in the $|pos\rangle,|neg\rangle$ basis to 
establish the synchronization with {\it A}.

At this point the authors note that the protocol is incomplete because 
$|pos\rangle$ and $|neg\rangle$ are not uniquely determined.  In the 
case of an atomic clock the $\pi/2$ pulse that defines {\it H} in that 
context has a phase relative to the atoms.  This is the relative phase 
between the interrogation oscillator and the precession of the system 
that prevents QCSP eq. 2 from representing a real atomic clock.
Since knowing this phase is equivalent to solving the problem in the 
first place (as the authors point out), they propose adding a second 
oscillation frequency and observing the ``beat note'' between the two 
frequencies to eliminate 
the phase ambiguity, $\delta$.  However for this to work, $\delta$ 
must be the same for both frequencies.   In general this would not be 
true because it must include the phase shift due to the transport 
process 
which will usually be frequency dependent. If 
the two $\delta$'s in QCSP 
eq. 7 are {\it not} the same, then the beat envelope is not 
independent of this additional phase ambiguity and it hasn't been 
resolved. 
For example, if the particles are different isotopes of an atom, 
then their sensitivities to external perturbations will be different 
leading to a different accumulated phase.  To ignore this is 
equivalent to
once again assuming a perfect ESCT.  In addition the two accumulated 
phases (for each species) are likely to be time-dependent once the 
trip is complete (imperfect clocks).

Since the arbitrary phase $\delta$ is not resolved, time 
synchronization is not achieved.  Even if the rest of QCS were to 
work, the best one could hope for would be frequency synchronization 
(``syntonization'') which is a much easier class of problem.  
However given our objections, as it 
stands, QCS probably isn't able to do syntonization without perfect 
clocks and as has already been pointed out, 
perfect clocks provide syntonization ``for free''. 
On the other hand, there may be ways to 
still get useful 
information.  
For instance, if the clocks belonging to {\it A} and {\it B} are known to differ 
only by a constant rate then {\it A} and {\it B} could double the size of their 
ensembles and perform two separate QCS protocols.  The difference in 
results would be the phase accumulated {\it only} while the ensembles 
were in their remote locations, and not due to the transport process 
(of course this necessarily includes phase errors introduced by 
imperfect environmental control, but these will be significantly less 
than those introduced by the transport process).  
This procedure, together with the assumption of a linear rate would give 
the frequency offset of the two clocks.  They could further increase 
the size of their ensembles (all transported at the same time) and 
make additional QCS measurements at later times to retain the 
frequency information over longer periods. The two clocks will still not 
be phase-locked, but important error-free information will have been determined 
about their respective operation at remote sites.

At first, it may seem that a way around the problems of transporting 
entangled ensembles over large distances may be overcome by 
using entangled photons which are sent from {\it A} to {\it B}. The 
entanglement is then transferred to the system of choice (e.g., atoms) 
for storage and carrying out of the protocol.  Leaving aside the 
engineering details of accomplishing this on a global scale, one must 
know the propagation delay of the photons 
in order to determine the phase of the entanglement that is 
transferred.  We are assuming that the phase of the photon is written 
onto the phase of the atom during entanglement transfer which is true 
in most resonant atom-photon interactions.
With this assumption, transferring the entanglement with photons 
is also still equivalent to the classical problem.
 
In summary we believe that the QCS protocol is an interesting area of 
study, but that as stated, it will work no better than existing 
classical methods.  We emphasize that our objections are not based on the 
existence of hard engineering problems (which there are), but that 
the problems that must be solved for QCS to work (perfect clock 
trips) appear to be the same as those for classical methods.  However, 
that is 
not to say that there isn't some other 
protocol that will make QCS work.

We gratefully acknowledge helpful discussions with D. Branning,
R. Hughes, B. Klipstein,
P. Kwiat, S. Lamoreaux, D. Matsakis, A. Migdall, T. Porto and A. White.


%
%

%
%

\end{document}